\pdfoutput=1
\documentclass{article}
\usepackage{spconf,amsmath,graphicx,multirow,makecell}    
\usepackage{subfigure}
\usepackage{hyperref}
\usepackage{verbatim}     
\usepackage{gensymb}
\usepackage{times}
\usepackage{amssymb}
\usepackage{bm}
\usepackage{soul}
\usepackage{url}
\usepackage{algorithmic}
\usepackage{tabularx}
\usepackage{xcolor}
\usepackage{stfloats}
\usepackage{caption}
\usepackage{siunitx}
\usepackage{flushend}
\usepackage{caption}
\usepackage{array}

\title{single image super-resolution via residual neuron attention networks}
\name{Wenjie Ai, Xiaoguang Tu$^{*}$, \thanks{Xiaoguang Tu is the corresponding author}Shilei Cheng, Mei Xie}
\address{School of Information and Communication Engineering, \\ University of Electronic Science and Technology of China, CN}
\begin{document}
	%
	\maketitle
	\begin{abstract}
		Deep Convolutional Neural Networks (DCNNs) have achieved impressive performance in Single Image Super-Resolution (SISR). To further improve the performance, existing CNN-based methods generally focus on designing deeper architecture of the network. However, we argue blindly increasing network's depth is not the most sensible way. In this paper, we propose a novel end-to-end Residual Neuron Attention Networks (RNAN) for more efficient and effective SISR. Structurally, our RNAN is a sequential integration of the well-designed Global Context-enhanced Residual Groups (GCRGs), which extracts super-resolved features from coarse to fine. Our GCRG is designed with two novelties. Firstly, the Residual Neuron Attention (RNA) mechanism is proposed in each block of GCRG to reveal the relevance of neurons for better feature representation. Furthermore, the Global Context (GC) block is embedded into RNAN at the end of each GCRG for effectively modeling the global contextual information.  Experiments results demonstrate that our RNAN achieves the comparable results with state-of-the-art methods in terms of both quantitative metrics and visual quality, however, with simplified network architecture.
		
		
	\end{abstract}
	\begin{figure*} [pt]
		\centering
		\includegraphics{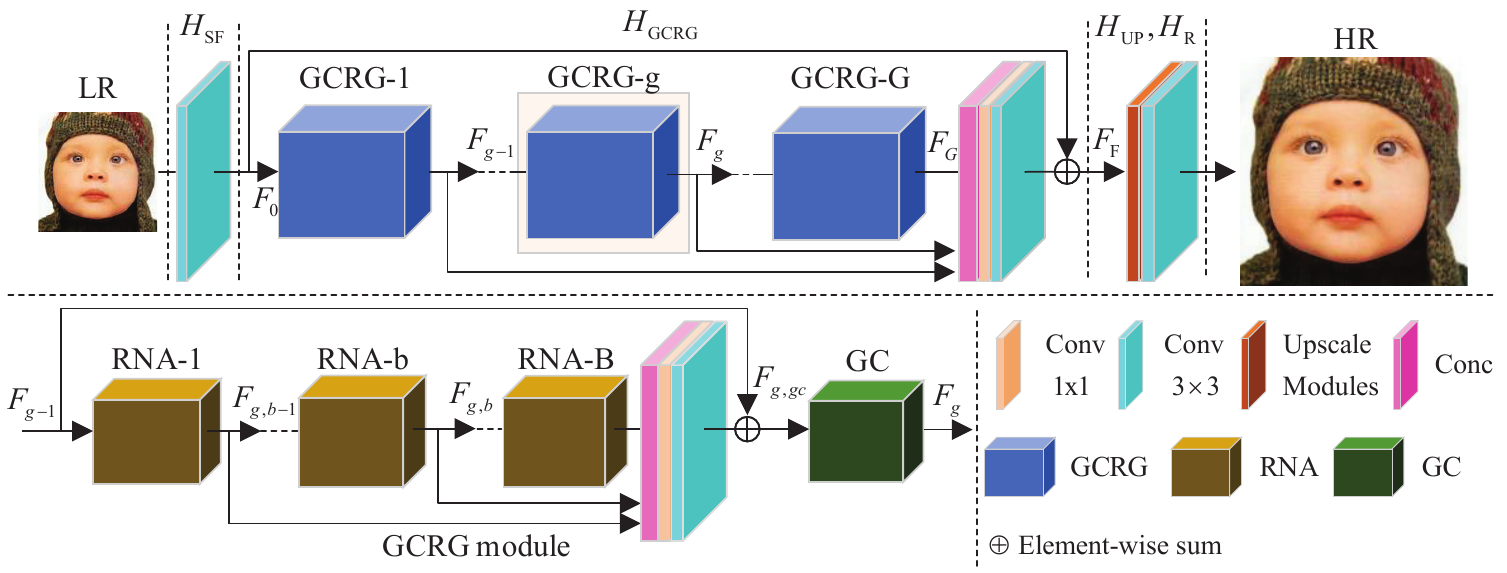}
		\setlength{\belowcaptionskip}{-15pt}
		\footnotesize
		\caption{Overview of RNAN. RNAN is a cascaded combination of the proposed Global Context-enhanced Residual Groups (GCRGs). As is shown in the upper panel, RNAN consists of four modules, the shallow feature extractor ($H_{\text{SF}}$), global context-enchanced residual groups ($H_{\text{GCRG}}$), up-sampling module ($H_{\text{UP}}$), and reconstruction layer ($H_{\text{R}}$). }
		\label{fig:net}
	\end{figure*}
	
	\section{Introduction}
	
	Single Image Super-Resolution (SISR) aims to reconstruct the visually High-Resolution (HR) images from the Low Resolution (LR) ones, which has various applications such as satellite imaging \cite{yildirim2012novel}, medical imaging \cite{shi2013cardiac,tu2017automatic,tu2016mr} and small object detection \cite{noh2019better,tu2020face}. However, given a specific LR image, the mapping to an HR one could have many solutions, making this task ill-posed. Benefiting from the powerful feature representation and end-to-end training, Convolutial Neural Networks (CNNs) have  demonstrated significant achievements in various computer vision tasks, greatly promoted the development of SISR. In the work  \cite{dong2014learning}, Dong \textit{et al}. firstly proposed SRCNN with three-layer to map a LR image to a Super-Resolution (SR) one. Later, networks are designed with deeper and complicated structure to further improve the performance. Deepening the networks has been considered useful in SISR methods, especially when He \textit{et al}. \cite{he2016deep} proposed ResNet with residual learning and Huang \textit{et al}. \cite{huang2017densely} raised DenseNet based on dense connections. Later, Lim \textit{et al}. \cite{lim2017enhanced} designed a very deep network termed as EDSR by stacking residual blocks for super resolution. Furthermore, Zhang \textit{et al}.\cite{zhang2018residual} combined both residual learning and dense connections to sufficiently utilize the hierarchical featWEures from different convolutional layers to further enhance the SR performance. The excellent performance has verified the importance of the depth representation for SISR. However, we argue that simply deepening the network is not the desired way for SISR as the relevance of features has not been thoroughly explored.
	
	To address the issues mentioned above, several CNN-based methods have been exploited, focusing on the attention of particular features for SISR. For example, Liu \textit{et al}.\cite{liu2018non} made use of the non-local attention block proposed in \cite{wang2018non} for image restoration. In \cite{li2019image}, Li \textit{et al}. utilized spatial attention module and DenseNet to reconstruct realistic HR images. Different from the methods as mentioned above that only exploit correlations in spatial space, other methods attempted to explore the channel correlation of features. In  \cite{zhang2018image}, Zhang \textit{et al}. utilized the channel attention block (SE) \cite{hu2018squeeze} to improve the performance of SR. Later, methods like \cite{kim2018ram, muqeet2019hybrid, tu2019learning, tu2019deep} made full utilization of both spatial attention and channel attention to improve the SR performance.
	
	Inspired by the above methods, we propose a novel Residual Neuron Attention Network (RNAN) for better representation and learning of features, as well as exploiting long-range global contextual information to enhance SISR. On the one hand, we propose the RNA blocks for explicitly modeling the interdependencies between the neurons of features, which is able to selectively re-weight the key neurons to learn more characteristic features. On the other hand, a global context block is embedded into GCRG to further model the correlations of global contextual information. The experimental results have shown that our method can effectively improve the quantitative results and visual quality compared with state-of-the-art methods.
	
	Our contributions are summarized as follows:
	
	\textbullet We elaborate the cascaded Global Context-enhanced Residual Groups (GCRGs) to construct a novel Residual Neuron Attention Networks (RNAN) for Single Image Super-Resolution (SISR).
	
	\textbullet We propose a Residual Neuron Attention (RNA) to concentrate more on neuron-wise relationships, as well as employing a lightweight Global Context (GC) block  at the end of each GCRG, to incorporate global contextual information.
	
	\textbullet Extensive experiments on several benchmark datasets demonstrate that our RNAN achieves superior results with fewer parameters.
	
	\vspace{-0.2cm}
	\section{Proposed method}
	
	
	\subsection{Network architecture}
	As shown in \autoref{fig:net}, our RNAN can be divided into four parts, \textit{i.e.,} shallow feature extractor, Global Context-enhanced Residual Groups (GCRGs), up-sampling module, and reconstruction layer. Given $ I_{\text{LR}} $ and $ I_{\text{HR}} $ as the input and output of RNAN, respectively. Following the work \cite{lim2017enhanced, zhao2019multi, tu2019enhance}, we apply only one convolutional layer to extract the shallow features $ F_{0} $ from the LR input
	\begin{equation}
	\small
	\setlength{\abovedisplayskip}{3pt}
	\setlength{\belowdisplayskip}{3pt}
	{F}_{0}=H_{\text{SF}}\left({I}_{\text{LR}}\right),
	\end{equation}
	where $ H_{\text{SF}} $ represents the convolutional operation to extract features from the shallow layers, $F_{0}$ is the input of GCRGs. Suppose we have \textit{G} GCRGs, the output $F_{g}$ of the \textit{g}-th GCRG can be expressed as
	\begin{equation}
	\begin{aligned}
	\small
	\setlength{\abovedisplayskip}{3pt}
	\setlength{\belowdisplayskip}{3pt}
	{F_{\text{g}}} &=H_{\text{GCRG,g}}\left(F_{\text{g-1}}\right) \\
	&=H_{\text{GCRG,g}}\left(H_{\text{GCRG,g-1}}\left(\cdots\left(H_{\text{GCRG,1}}\left(F_{0}\right)\right) \cdots\right)\right)
	\end{aligned}
	\end{equation}
	where $H_{\text{GCRG,g}}$ denotes the representation of \textit{g}-th GCRG. The GCRG is used to enhance the sensitivity of feature maps, as well as capturing global contextual information. Then we extract features from each GCRG block, and conduct uniform-spaced features fusion. To stabilize the training, we introduce a global residual learning as
	\begin{equation}
	\setlength{\abovedisplayskip}{3pt}
	\setlength{\belowdisplayskip}{3pt}
	{F}_{\text{F}} = {F}_{0} + H_{\text{RF}}\left(H_{\text{F}}\left(F_{\text{N}},F_{\text{2N}}, ..., F_{\text{G-N}}, F_{\text{G}}\right)\right)
	\end{equation}
	where $F_{\text{F}}$ is the output features of GCRGs, $H_{\text{F}}$ represents feature fusion which concatenates the outputs of uniformly-spaced GCRGs with an interval $\text{N}$ (\textit{e.g}., $\text{N}$ equals 2), and $H_{\text{RF}}$ denotes the convolutional layers, including a $1\times1$ convolutional (conv) layer for feature dimension reduction and a $3\times3$ conv layer for further feature fusion. 
	After that, the up-sampling module upsamples the residual learned feature maps $F_{\text{F}}$, followed by reconstruction layer
	\begin{equation}
	\setlength{\abovedisplayskip}{3pt}
	\setlength{\belowdisplayskip}{3pt}
	I_{\text{SR}} = H_{\text{R}}\left(H_{\text{UP}}\left(F_{\text{F}}\right)\right) = H_{\text{RANR}}\left(I_{\text{LR}}\right)
	\end{equation}
	where $H_{\text{R}}$ and $H_{\text{UP}}$ denote the reconstruction layer and upsampling module, respectively. $H_{\text{RANR}}$ is the representation of the proposed RNAN. Inspired by the work \cite{shi2016real}, we use sub-pixel convolutional layer as our up-sampling module. The reconstruction layer employs three $ 3\times3$ convolutional kernels to generate the 3-channel  super-resolved RGB image. It is worth noting that using residual learning and concatenation in global architecture and every GCRG can bypass more abundant low-frequency information during training \cite{lim2017enhanced, zhang2018residual}.

	\begin{figure}[t]
		\centering
		\includegraphics{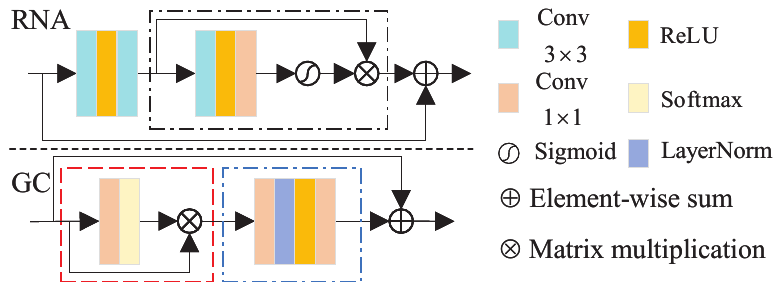}
		\setlength{\belowcaptionskip}{0pt}
		\footnotesize
		\caption{Upper: illustration of the Residual Neuron Attention (RNA) block, black-dashed rectangle shown in figure denotes Neuron Attention (NA) mechanism. Bottom: illustration of the Global Context (GC) block, red-dashed rectangle and blue-dashed rectangle stand for context modeling and feature transform, respectively.}
		\label{fig:sub}
		\vspace{-2em}
	\end{figure}
	\vspace{-0.2cm}
	\subsection{Global Context-enhanced Residual Group}
	We now give more details for the proposed GCRG, which is composed of several (10 in our experiments) Residual Neuron Attention (RNA) blocks and one Global Context (GC) block. In order to further facilitate feature extraction, we uniformly-spaced concatenate the hierarchical features that generated from RNAs, the same with feature fusion of different GCRG blocks. Therefore, the final representation of the g-th GCRG can be defined as
	\begin{equation}
	\small
	\setlength{\abovedisplayskip}{3pt}
	\setlength{\belowdisplayskip}{3pt}
	F_{\text{g}} = H_{\text{GC}}\left(F_{\text{g-1}}+H_{\text{DF}}\left(H_{\text{F}}\left(F_{\text{g,M}}, F_{\text{g,2M}},...,F_{\text{G-M}}, F_{\text{G}}\right)\right)\right)
	\end{equation}
	where $F_{\text{g}}$ and $F_{\text{g-1}}$ are the output and input of the \textit{g-}th GCRG, respectively. $H_{\text{F}}$ denotes feature concatenation, and $H_{\text{DF}}$ denotes convolutions with the kernel size as $1\times1$ and $3\times3$, respectively. $\text{M}$ denotes the interval that we concatenate the features of RNA blocks.
	
	
	\vspace{-0.2cm}
	\subsubsection{Residual Neuron Attention (RNA) block}
	
	Inspired by the Residual Blocks (RB) in \cite{wang2017residual, he2016deep, tu2019joint} and the Neuron Attention (NA) in \cite{qin2019nasnet}, we integrate NA into RB and propose Residual Neuron Attention (RNA) block, as shown in \autoref{fig:sub}. 
	Taking the input and output features of the \textit{b-}th RNA in \textit{g-}th GCRG as $F_{\text{g,b-1}}$, and $F_{\text{g,b}}$, respectively, the process of RNA can be formulated as
	\begin{equation}
	\small
	\setlength{\abovedisplayskip}{3pt}
	\setlength{\belowdisplayskip}{3pt}
	F_{\text{g,b}} = F_{\text{g,b-1}}+F_{\text{NA}}\left(F_{\text{RB}}\left(F_{\text{g,b-1}}\right)\right)
	\end{equation}
	where $F_{\text{NA}}$ and $F_{\text{RB}}$ denote NA module and RB, respectively.
	
	Previous CNN-based methods utilize convolutional filters to incorporate channel-wise and spatial-wise information within local receptive field to generate the final convolutional feature.
	However, the contextual information outside the local receptive field in the last convolutional layer can not be used. To this end, we exploit the independencies of neurons modeled by Neuron Attention (NA) mechanism to recalibrate neuron-wise responses adaptively and dynamically. NA consists of two main operations, Depthwise Convolution (DC) and Pointwise Convolution (PC).
	DC aims to make use of spatial information in each individual channel, which keeps the number of filters the same with channels of input features. To overcome the drawback of DC that can not fully utilize the information of different maps in the same spatial location, we adopt the PC, using $1\times1$ convolution kernel with the number of filters the same with the depth of input features. Similar with the attention mechanism in \cite{wang2017residual}, we employ a sigmoid activation function after the PC. The operations of NA can be expressed as
	\begin{equation}
	\setlength{\abovedisplayskip}{3pt}
	\setlength{\belowdisplayskip}{3pt}
	Y=X\otimes\left(\sigma\left(W_{p}\left(\delta\left(W_{d}\left(X\right)\right)\right)\right)\right)
	\end{equation}
	where $W_{d}$ and $ W_{p} $ denote the weight of the DC and the PC, respectively. $\sigma$ and $\delta$ represent the sigmoid and ReLU activation function, respectively. \textit{X} is the input features, and \textit{Y} is the corresponding output. With the NA module, the residual component in RNA can be adaptively recalibrated.
	\vspace{-0.2cm}
	\subsubsection{Global Context (GC) module.}
	The Global Context (GC) block \cite{cao2019gcnet} is placed at the end of each GCRG to learn global contextual information. GC mainly consists of context modeling and feature transform, as illustrated in the bottom panel of \autoref{fig:sub}. In this way, GC can benefit model learning by both the simplified non-local block and the Squeeze-Excitation (SE) block \cite{hu2018squeeze}. The former can effectively model long-range dependencies throughout the full image with smaller computation cost compared with original non-local block \cite{wang2018non}. Meanwhile, the latter can fully capture channel-wise dependencies. 
	
	We denote $F_{g,gc}=\left\{{\mathbf{x}_{i}}\right\}_{i=1}^{N_{p}}$ as the fused feature maps of multiple RNA blocks; $F_{g}=\left\{{\mathbf{z}_{i}}\right\}_{i=1}^{N_{p}}$ as the output of GCRG, where $N_{p}$ is the number of positions in the feature map (\textit{e.g}., $N_{p}=H\times{W}$ in an image). The detailed architecture of the GC block is illustrated in the bottom panel of \autoref{fig:sub}. GC block can be formulated as
	\begin{equation}
	\setlength{\abovedisplayskip}{3pt}
	\setlength{\belowdisplayskip}{3pt}
	\mathbf{z}_{i}=\mathbf{x}_{i}+{W}_{v2}{\delta}\left(LN\left(W_{v1}\sum_{j=1}^{N_{p}}\frac{e^{W_{k}\mathbf{x}_{j}}}{\sum_{m=1}^{N_{p}}e^{W_{k}\mathbf{x}_{m}}}\right)\right)
	\end{equation}
	where $W$ denotes convolution operation,  $W_{v2}\delta\left(LN\left(W_{v1}\left(\cdot \right)\right)\right)$  denotes the features bottleneck transform, and $\left(\cdot\right)$ denotes the global context modeling. $\delta$ and $LN$ stand for ReLU and LayerNorm, respectively. We set the bottle ratio $\textit{r}$ as 16 in our experiments.
	
	\begin{table*}[t]
		\footnotesize
		\centering
		\caption{Results of various SR methods.The best and second best values are highlighted in blod and underline in italic. Results using self-ensemble were denoted with +.}
		\label{tab:results}
		\begin{tabular}{|c|c|c|c|c|c|c|c|c|c|c|c|c|}
			\hline
			& \multicolumn{3}{c|}{Set5 (PSNR/SSIM)} & \multicolumn{3}{c|}{Set14 (PSNR/SSIM)} & \multicolumn{3}{c|}{BSD100 (PSNR/SSIM)} & \multicolumn{3}{c|}{Urban100 (PSNR/SSIM)}   \\ \hline
			
			Methods     & $\times2$     & $\times3 $    & $\times4$
			& $\times2$     & $\times3 $    & $\times4$
			& $\times2$     & $\times3 $    & $\times4$
			& $\times2$     & $\times3 $    & $\times4$      \\ \hline
			
			\multirow{2}*{Bicubic} &
			33.66/      & 30.39/   & 28.42/
			& 30.24/   & 27.55/   & 26.00/
			& 29.56/   & 27.21/   & 25.96/
			& 26.88/   & 24.46/   & 23.14/ \\
			
			& .9299   & .8682   & .8104
			& .8688   & .7742   & .7027
			& .8431   & .7385   & .6675
			& .8403   & .7349   & .6577   \\ \hline
			
			\multirow{2}*{SRCNN} &
			36.66/    & 32.75/   & 30.48/
			& 32.45/    & 29.30/   & 27.50/
			& 31.36/    & 28.41/   & 26.90/
			& 29.50/    & 26.24/   & 24.52/   \\
			
			& .9542  & .9090  & .8628
			& .9067  & .8215  & .7513
			& .8879  & .7863  & .7101
			& .8946  & .7989  & .7221  \\ \hline
			
			\multirow{2}*{FSRCNN} &
			37.05/   & 33.18/  & 30.72/
			& 32.66/   & 29.37/  & 27.61/
			& 31.53/   & 28.53/  & 26.98/
			& 29.88/   & 26.43/  & 24.62/  \\
			
			& .9560  & .9140 & .8660
			& .9090  & .8240 & .7550
			& .8920  & .7910 & .7150
			& .9020  & .8080 & .7280  \\ \hline
			
			\multirow{2}*{VDSR} &
			37.53/  & 33.67/  & 31.35/
			& 33.05/   & 29.78/  & 28.02/
			& 31.90/   & 28.83/  & 27.29/
			& 30.77/   & 27.14/  & 25.18/   \\
			
			& .9590  & .9210  & .8830
			& .9130  & .8320  & .7680
			& .8960  & .7990  & .0726
			& .9140  & .8290  & .7540   \\ \hline
			
			\multirow{2}*{LapSRN} &
			37.52/   & 33.82/  & 31.54/
			& 33.08/  & 29.87/  & 28.19/
			& 31.08/  & 28.82/  & 27.32/
			& 30.41/  & 27.07/  &25.21/  \\
			
			& .9591   & .9227  & .8850
			& .9130   & .8320  & .7720
			& .8950   & .7980  & .7270
			& .9101   & .8280  & .7560 \\ \hline
			
			\multirow{2}*{NLRN} &
			38.00/   & 34.27/  & 31.92/
			& 33.46/ & 30.16/  & 28.36/
			& 32.19/ & 29.06/  & 27.48/
			& 31.81/ & 29.06/  & 25.79/      \\
			
			& .9603  & .9266  & .8916
			& .9159  & .8374  & .7745
			& .8992  & .8026  & .7346
			& .9246  & .8453  & .7729        \\ \hline
			
			\multirow{2}*{EDSR} &
			38.11/  & 34.65/  & 32.46/
			& 33.92/ & 30.52/   & 28.80/
			& 32.32/ & 29.25/  & 27.71/
			& 32.93/ & 28.80/  & 26.64/         \\
			
			& .9602  & .9280  & .8968
			& .9195  & .8462  & .7876
			& .9013  & .8093  & .7420
			& .9351  & .8653  & .8033   \\ \hline
			
			\multirow{2}*{RDN} &
			38.24/   & 34.71/  & 32.47/
			& \underline{34.01}/ & 30.57/  & 28.81/
			& 32.34/ & 29.26/  & 27.72/
			& 32.89/ & 28.80/  & 26.61/      \\
			
			& .9614  & .9296  & \underline{.8990}
			& \underline{.9212}  & .8468  & .7871
			& .9017  & .8093  & .7419
			& .9353  & .8653  & .8028      \\
			\Xhline{1pt}
			\multirow{2}*{RNAN} &
			\underline{38.24}/ & \underline{34.73}/  & \underline{32.52}/
			& 33.97/ & \underline{30.59}/ & \underline{28.82}/
			& \underline{33.07}/ & \underline{29.26}/ & \underline{27.72}/
			& \underline{33.07}/ & \underline{28.85}/ & \underline{26.67}/    \\
			
			& \underline{.9614} & \underline{.9297} & .8986
			& .9211 & \underline{.8473} & \underline{.7872}
			& \underline{.9021} & \underline{.8096} & \underline{.7418}
			& \underline{.9368} & \underline{.8667} & \underline{.8049}     \\ \hline
			
			\multirow{2}*{RNAN+} &
			\textbf{38.31} / & \textbf{34.80} /  &  \textbf{32.66}/
			& \textbf{34.10}/ & \textbf{30.69}/ & \textbf{28.92}/
			& \textbf{33.42}/ & \textbf{29.33}/ &  \textbf{27.79}/
			& \textbf{33.28}/ & \textbf{29.08}/ &  \textbf{26.90}/  \\
			
			& \textbf{.9617} & \textbf{.9302} & \textbf{.9005}
			& \textbf{.9221} & \textbf{.8486} & \textbf{.7894}
			& \textbf{.9027} & \textbf{.8108} & \textbf{.7432}
			& \textbf{.9384} & \textbf{.8699} & \textbf{.8097} \\ \hline
			
			
		\end{tabular}
		\vspace{-1em}
		
	\end{table*}
	\vspace{2pt}
	\vspace{-0.2cm}
	\section{Experiments}
	\vspace{-0.2cm}
	\subsection{Setting}
	Following \cite{kim2016accurate,lim2017enhanced}, we use 800 images from DIV2K datasets as the training set. The LR images were obtained by bicubic downsampling of HR images using MATLAB. For testing, we use four standard benchmark datasets:Set5, Set14, B100, and Urban100.

	During training, we randomly cropped patches from LR images and corresponding HR images. Besides, we augment the training images by randomly rotating 90$^{\circ}$, 180$^{\circ}$, 270$^{\circ}$ and horizontally flipping. In every training mini-batch, 16 cropped and colorful LR patches with size of $48\times48$ are provided as inputs.We train our model with Adam optimizer with $\beta_{1}=0.9$, $\beta_{2}=0.999$, and $L_1$ to calculate the loss between input and output. The initial learning rate is assigned by 0.0001, and decreases to half every 200 epochs. Moreover, we set the numbers of RNAB as 20 and GCRG as 10.
	Similar to \cite{lim2017enhanced}, self-ensemble, that averages the outputs of augmented inputs of one image when testing,  was introduced to maximize the potential performance of our model.
	
	\begin{figure}[h]
		\centering
		\footnotesize
		\subfigure{
			\hspace{0.2cm}
			\centering
			\begin{minipage}[ht]{0.18\linewidth}
				\vspace{0.5pt}
				\centerline{\includegraphics[width=2.1cm, height=2.5cm]{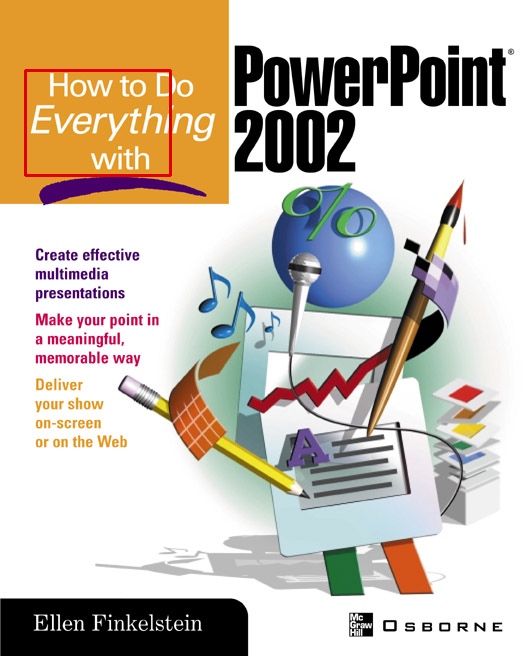}}
				\centerline{ppt3 (Set14)}\smallskip
				\vspace{1.5pt}
				\centerline{\includegraphics[width=2.1cm, height=2.5cm]{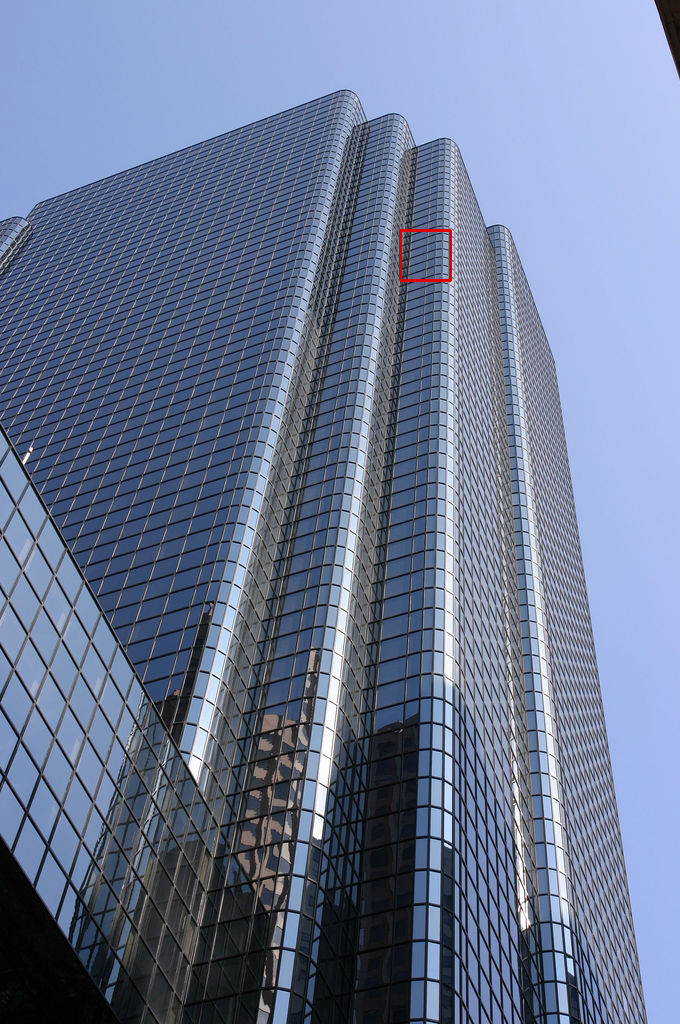}}
				\centerline{img074 (Urban100)}\smallskip
			\end{minipage}
			\hspace{0.3cm}
			\begin{minipage}[ht]{0.18\linewidth}
				\centerline{\includegraphics[width=1.4cm,height=1.0cm]{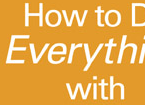}}
				\centerline{HR}\smallskip
				\centerline{\includegraphics[width=1.4cm,height=1.0cm]{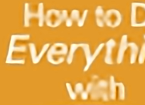}}
				\centerline{VDSR}\smallskip
				\vspace{3pt}
				\centerline{\includegraphics[width=1.4cm,height=1.0cm]{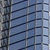}}
				\centerline{HR}\smallskip
				\centerline{\includegraphics[width=1.4cm,height=1.0cm]{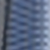}}
				\centerline{VDSR}\smallskip
			\end{minipage}
			\hfill
			\begin{minipage}[ht]{0.18\linewidth}
				\vspace{1.5pt}
				\centerline{\includegraphics[width=1.4cm,height=1.0cm]{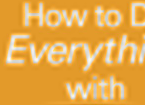}}
				\centerline{Bicubic}\smallskip
				\centerline{\includegraphics[width=1.4cm,height=1.0cm]{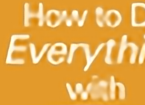}}
				\centerline{LapSRN}\smallskip
				\vspace{1.5pt}
				\centerline{\includegraphics[width=1.4cm,height=1.0cm]{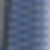}}
				\centerline{Bicubic}\smallskip
				\centerline{\includegraphics[width=1.4cm,height=1.0cm]{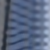}}
				\centerline{LapSRN}\smallskip
			\end{minipage}
			\hfill
			\begin{minipage}[ht]{0.18\linewidth}
				\centerline{\includegraphics[width=1.4cm,height=1.0cm]{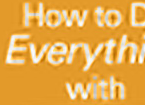}}
				\centerline{SRCNN}\smallskip
				\centerline{\includegraphics[width=1.4cm,height=1.0cm]{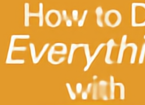}}
				\centerline{EDSR}\smallskip
				\vspace{3pt}
				\centerline{\includegraphics[width=1.4cm,height=1.0cm]{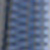}}
				\centerline{SRCNN}\smallskip
				\centerline{\includegraphics[width=1.4cm,height=1.0cm]{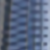}}
				\centerline{EDSR}\smallskip
			\end{minipage}
			\hfill
			\begin{minipage}[ht]{0.18\linewidth}
				\centerline{\includegraphics[width=1.4cm,height=1.0cm]{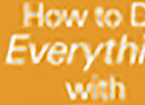}}
				\centerline{FSRCNN}\smallskip
				\centerline{\includegraphics[width=1.4cm,height=1.0cm]{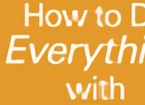}}
				\centerline{RNAN}\smallskip
				\vspace{3pt}
				\centerline{\includegraphics[width=1.4cm,height=1.0cm]{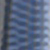}}
				\centerline{FSRCNN}\smallskip
				\centerline{\includegraphics[width=1.4cm,height=1.0cm]{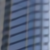}}
				\centerline{RNAN}\smallskip
			\end{minipage}
			\hfill	
		}
		\setlength{\abovecaptionskip}{-6pt}
		\caption{Visual comparison for $4\times$ SR on datasets of Set14 and Urban100. Please zoom in for better visualization. }
		\label{fig:res}
		\vspace{-2em}
	\end{figure}
	\vspace{-0.2cm}
	\subsection{Comparison with state-of-the-art methods}
	We compare our RNAN with several state-of-the-art SR methods: SRCNN \cite{dong2014learning}, FSRCNN \cite{dong2016accelerating},
	VDSR \cite{kim2016accurate}, LapSRN \cite{lai2017deep}, EDSR \cite{lim2017enhanced},
	NLRN \cite{liu2018non}, RDN \cite{zhang2018residual}. The performance of different models are executed with quantitative and qualitative comparisons.
	
	For fair comparison, we follow a common setting \cite{lim2017enhanced, huang2017densely, liu2018non}, evaluating our model using the luminance channel (Y) of the transformed YCbCr space for quantitative measurement. \autoref{tab:results} shows the quantitative results of PSNR and SSIM values of the compared SR methods for $\times2$, $\times3$, and $\times4$ super resolution, respectively. Referring to \autoref{tab:results}, RNAN$+$ which adopts self-ensemble strategy, achieves better performance on all benchmark datasets regarding various scaling factors, compared with other methods. Without self-ensemble, RNAN and RDN achieve vary similar results and still outperform other methods, however RNAN has less parameters than that of RDN (about $4/5$, see \autoref{tab:params}). Besides, we observe that the gap between RNAN and EDSR decreases as the upsampling factor increases (\textit{e.g.}, $\times2$: 0.13dB, $\times3$: 0.08dB, $\times4$: 0.04dB in Set14), but the slightly better performance of RNAN on scale $\times4$ brings about significantly visual advance (see \autoref{fig:res}). It is worth to note that the parameters of RNAN are about $2/5$ of EDSR. \autoref{tab:results} and \autoref{tab:params} show that our proposed models increase the performance with better trade-off between parameters and performance.
	\begin{table}[t]
		\centering
		\footnotesize
		\setlength{\belowcaptionskip}{5pt}
		\caption{The number of parameters of RNAN and other SR methods (Unit: M). RNAN and RNAN$+$ have the similar number of parameters on different scales.}
		\renewcommand{\arraystretch}{1.1}
		\begin{tabular}{c|c|c|c|c|c|c}
			\hline
			\small{method} & FSRCNN & LapSRN & VDSR & EDSR  & RDN   & RNAN \\ \hline
			\small{Params} & 0.01 & 0.81  & 0.67  & 43.10 & 22.27 & 17.30\\ \hline
		\end{tabular}
		\label{tab:params}
		\vspace{-2em}
	\end{table}
	
	In \autoref{fig:res}, we visually illustrate the qualitative comparisons on scale $\times 4$ on images from Set14 and Urban100. It is clear to see that RNAN recovers more details than the compared SR methods. For the image 'ppt3' from Set14 dataset, RNAN can generate more clearly distinguishable words than other methods. Referring to the image 'img074' from Urban100 dataset, the compared methods cannot reconstruct the realistic and clear structure of the building. On the contrary, RNAN reconstructs the image that is more faithful to the ground truth with sharper edges and more high-frequency details. Such obvious comparisons demonstrate that networks with NA and GC can extract more sophisticated features from the LR image.
	
	\vspace{-0.3cm}
	\section{Conclusion}
	In this paper, we propose a Residual Neuron Attention Networks (RNAN) for high-realistic image super resolution. Specifically, we propose the Global Context-enhanced Residual Groups (GCRGs), each composed of multiple Residual Neuron Attention (RNA) blocks and one Global Context (GC) block, to recalibrade neuron-wise feature responses adaptively and capture global contextual information. Extensive experiments on several benchmark datasets demonstrate that our RNAN can significantly improve the super resolution performance with fewer parameters involved.
	
	\bibliographystyle{IEEEbib}
	\bibliography{strings,refs}

\begin{thebibliography}{10}

\bibitem{yildirim2012novel}
Deniz Y{\i}ld{\i}r{\i}m and O{\u{g}}uz G{\"u}ng{\"o}r,
\newblock ``A novel image fusion method using ikonos satellite images,''
\newblock {\em GGS}, vol. 1, no. 1, pp. 75--83, 2012.

\bibitem{shi2013cardiac}
Wenzhe Shi, Jose Caballero, Christian Ledig, Xiahai Zhuang, Wenjia Bai, Kanwal
  Bhatia, Antonio M Simoes~Monteiro de~Marvao, Tim Dawes, Declan O’Regan, and
  Daniel Rueckert,
\newblock ``Cardiac image super-resolution with global correspondence using
  multi-atlas patchmatch,''
\newblock in {\em MICCAI}. Springer, 2013, pp. 9--16.

\bibitem{tu2017automatic}
Xiaoguang Tu, Mei Xie, Jingjing Gao, Zheng Ma, Daiqiang Chen, Qingfeng Wang,
  Samuel~G Finlayson, Yangming Ou, and Jie-Zhi Cheng,
\newblock ``Automatic categorization and scoring of solid, part-solid and
  non-solid pulmonary nodules in ct images with convolutional neural network,''
\newblock {\em Scientific reports}, vol. 7, no. 1, pp. 1--10, 2017.

\bibitem{tu2016mr}
Xiaoguang Tu, Jingjing Gao, Chongjing Zhu, Jie-Zhi Cheng, Zheng Ma, Xin Dai,
  and Mei Xie,
\newblock ``Mr image segmentation and bias field estimation based on coherent
  local intensity clustering with total variation regularization,''
\newblock {\em Medical \& biological engineering \& computing}, vol. 54, no.
  12, pp. 1807--1818, 2016.

\bibitem{noh2019better}
Junhyug Noh, Wonho Bae, Wonhee Lee, Jinhwan Seo, and Gunhee Kim,
\newblock ``Better to follow, follow to be better: Towards precise supervision
  of feature super-resolution for small object detection,''
\newblock in {\em CVPR}, 2019, pp. 9725--9734.

\bibitem{tu2020face}
XG~Tu, Y~Luo, HS~Zhang, WJ~Ai, Z~Ma, and M~Xie,
\newblock ``Face attribute invertion,''
\newblock {\em arXiv preprint arXiv:2001.04665}, 2020.

\bibitem{dong2014learning}
Chao Dong, Chen~Change Loy, Kaiming He, and Xiaoou Tang,
\newblock ``Learning a deep convolutional network for image super-resolution,''
\newblock in {\em ECCV}. Springer, 2014, pp. 184--199.

\bibitem{he2016deep}
Kaiming He, Xiangyu Zhang, Shaoqing Ren, and Jian Sun,
\newblock ``Deep residual learning for image recognition,''
\newblock in {\em CVPR}, 2016, pp. 770--778.

\bibitem{huang2017densely}
Gao Huang, Zhuang Liu, Laurens Van Der~Maaten, and Kilian~Q Weinberger,
\newblock ``Densely connected convolutional networks,''
\newblock in {\em CVPR}, 2017, pp. 4700--4708.

\bibitem{lim2017enhanced}
Bee Lim, Sanghyun Son, Heewon Kim, Seungjun Nah, and Kyoung Mu~Lee,
\newblock ``Enhanced deep residual networks for single image
  super-resolution,''
\newblock in {\em CVPRW}, 2017, pp. 136--144.

\bibitem{zhang2018residual}
Yulun Zhang, Yapeng Tian, Yu~Kong, Bineng Zhong, and Yun Fu,
\newblock ``Residual dense network for image super-resolution,''
\newblock in {\em CVPR}, 2018, pp. 2472--2481.

\bibitem{liu2018non}
Ding Liu, Bihan Wen, Yuchen Fan, Chen~Change Loy, and Thomas~S Huang,
\newblock ``Non-local recurrent network for image restoration,''
\newblock in {\em NIPS}, 2018, pp. 1673--1682.

\bibitem{wang2018non}
Xiaolong Wang, Ross Girshick, Abhinav Gupta, and Kaiming He,
\newblock ``Non-local neural networks,''
\newblock in {\em CVPR}, 2018, pp. 7794--7803.

\bibitem{li2019image}
Zhuangzi Li,
\newblock ``Image super-resolution using attention based densenet with residual
  deconvolution,''
\newblock {\em arXiv preprint arXiv:1907.05282}, 2019.

\bibitem{zhang2018image}
Yulun Zhang, Kunpeng Li, Kai Li, Lichen Wang, Bineng Zhong, and Yun Fu,
\newblock ``Image super-resolution using very deep residual channel attention
  networks,''
\newblock in {\em ECCV}, 2018, pp. 286--301.

\bibitem{hu2018squeeze}
Jie Hu, Li~Shen, and Gang Sun,
\newblock ``Squeeze-and-excitation networks,''
\newblock in {\em CVPR}, 2018, pp. 7132--7141.

\bibitem{kim2018ram}
Jun-Hyuk Kim, Jun-Ho Choi, Manri Cheon, and Jong-Seok Lee,
\newblock ``Ram: Residual attention module for single image super-resolution,''
\newblock {\em arXiv preprint arXiv:1811.12043}, 2018.

\bibitem{muqeet2019hybrid}
Abdul Muqeet, Md~Tauhid~Bin Iqbal, and Sung-Ho Bae,
\newblock ``Hybrid residual attention network for single image super
  resolution,''
\newblock {\em arXiv preprint arXiv:1907.05514}, 2019.

\bibitem{tu2019learning}
Xiaoguang Tu, Jian Zhao, Mei Xie, Guodong Du, Hengsheng Zhang, Jianshu Li,
  Zheng Ma, and Jiashi Feng,
\newblock ``Learning generalizable and identity-discriminative representations
  for face anti-spoofing,''
\newblock {\em arXiv preprint arXiv:1901.05602}, 2019.

\bibitem{tu2019deep}
Xiaoguang Tu, Hengsheng Zhang, Mei Xie, Yao Luo, Yuefei Zhang, and Zheng Ma,
\newblock ``Deep transfer across domains for face antispoofing,''
\newblock {\em Journal of Electronic Imaging}, vol. 28, no. 4, pp. 043001,
  2019.

\bibitem{zhao2019multi}
Jian Zhao, Jianshu Li, Xiaoguang Tu, Fang Zhao, Yuan Xin, Junliang Xing,
  Hengzhu Liu, Shuicheng Yan, and Jiashi Feng,
\newblock ``Multi-prototype networks for unconstrained set-based face
  recognition,''
\newblock {\em arXiv preprint arXiv:1902.04755}, 2019.

\bibitem{tu2019enhance}
Xiaoguang Tu, Hengsheng Zhang, Mei Xie, Yao Luo, Yuefei Zhang, and Zheng Ma,
\newblock ``Enhance the motion cues for face anti-spoofing using cnn-lstm
  architecture,''
\newblock {\em arXiv preprint arXiv:1901.05635}, 2019.

\bibitem{shi2016real}
Wenzhe Shi, Jose Caballero, Ferenc Husz{\'a}r, Johannes Totz, Andrew~P Aitken,
  Rob Bishop, Daniel Rueckert, and Zehan Wang,
\newblock ``Real-time single image and video super-resolution using an
  efficient sub-pixel convolutional neural network,''
\newblock in {\em CVPR}, 2016, pp. 1874--1883.

\bibitem{wang2017residual}
Fei Wang, Mengqing Jiang, Chen Qian, Shuo Yang, Cheng Li, Honggang Zhang,
  Xiaogang Wang, and Xiaoou Tang,
\newblock ``Residual attention network for image classification,''
\newblock in {\em CVPR}, 2017, pp. 3156--3164.

\bibitem{tu2019joint}
Xiaoguang Tu, Jian Zhao, Zihang Jiang, Yao Luo, Mei Xie, Yang Zhao, Linxiao He,
  Zheng Ma, and Jiashi Feng,
\newblock ``Joint 3d face reconstruction and dense face alignment from a single
  image with 2d-assisted self-supervised learning,''
\newblock {\em arXiv preprint arXiv:1903.09359}, 2019.

\bibitem{qin2019nasnet}
Xu~Qin and Zhilin Wang,
\newblock ``Nasnet: A neuron attention stage-by-stage net for single image
  deraining,''
\newblock {\em arXiv preprint arXiv:1912.03151}, 2019.

\bibitem{cao2019gcnet}
Yue Cao, Jiarui Xu, Stephen Lin, Fangyun Wei, and Han Hu,
\newblock ``Gcnet: Non-local networks meet squeeze-excitation networks and
  beyond,''
\newblock {\em arXiv preprint arXiv:1904.11492}, 2019.

\bibitem{kim2016accurate}
Jiwon Kim, Jung Kwon~Lee, and Kyoung Mu~Lee,
\newblock ``Accurate image super-resolution using very deep convolutional
  networks,''
\newblock in {\em CVPR}, 2016, pp. 1646--1654.

\bibitem{dong2016accelerating}
Chao Dong, Chen~Change Loy, and Xiaoou Tang,
\newblock ``Accelerating the super-resolution convolutional neural network,''
\newblock in {\em ECCV}. Springer, 2016, pp. 391--407.

\bibitem{lai2017deep}
Wei-Sheng Lai, Jia-Bin Huang, Narendra Ahuja, and Ming-Hsuan Yang,
\newblock ``Deep laplacian pyramid networks for fast and accurate
  super-resolution,''
\newblock in {\em CVPR}, 2017, pp. 624--632.

\end{thebibliography}
	
\end{document}